\documentclass[namedreferences]{solarphysics}

\usepackage[optionalrh]{spr-sola-addons} 
\usepackage{graphicx}        
\usepackage{color}           

\begin{document}
\begin{article}
\begin{opening}

\title{Evolution of Photospheric Magnetic Field and Electric Currents during the X1.6 Flare in Active Region NOAA 12192} 

\author[addressref={aff1},email={parthares@gmail.com}]{\inits{P.}\fnm{Partha}~\lnm{Chowdhury}\orcid{}} 
\author[addressref={aff2},corref,email={ravindra@iiap.res.in}]{\inits{B.}\fnm{Belur}~\lnm{Ravindra}\orcid{0000-0003-2165-3388}} 
\author[addressref={aff4},email={tiwari@baeri.org}]{\inits{S.}\fnm{Sanjiv Kumar}~\lnm{Tiwari}\orcid{}}

\address[id=aff1]{University College of Science and Technology, Chemical Technology Dept., University of Calcutta, 92, Acharya Prafulla Chandra Road, 700 009, Kolkata, India} 
\address[id=aff2]{Indian Institute of Astrophysics, II Block, Koramangala, Bengaluru – 560 034, India} 
\address[id=aff4]{Lockheed Martin Solar \& Astrophysics Laboratory, 3251 Hanover St. Bldg. 203, Palo Alto, CA 94306 \& Bay Area Environmental Research Institute, NASA Research Park, Moffett Field, CA 94035} 

\runningauthor{Chowdhury et~al.}
\runningtitle{\textit{magnetic fields and currents}}

\begin{abstract}
The dynamics of magnetic fields in the Sun's active regions plays a key role in triggering solar eruptions. Studies have shown that changes in the photosphere's magnetic field can destabilize large-scale structure of the corona, leading to explosive events such as flares and coronal mass ejections (CMEs). This paper delves into the magnetic field evolution associated with a powerful X1.6 class flare that erupted on October 22nd, 2014, from the flare-rich active region NOAA 12192. We utilized high-resolution vector magnetograms from the Helioseismic and Magnetic Imager (HMI) on NASA's Solar Dynamic Observatory (SDO) to track these changes. Our analysis reveals that a brightening, a precursor to the flare, began near the newly emerged, small-scale bipolar flux regions. During the X1.6 flare, the magnetic flux in both polarities displayed emergence and cancellation. The total current within the active region peaked during the flare. But, it is a non CME event and the ratio of direct to return current value remain close to 1. The large flare in this active region occured when the net current in both polarities attain the same sign. This implies that the Lorentz force, a consequence of the interaction between currents and magnetic fields, would have pushed the field lines together in this scenario. This reconnection of opposing magnetic fields is believed to be the driving force behind major flare occurred in this active region.
\end{abstract}


\end{opening}

\section{Introduction}
Sunspots represent regions on the surface of the Sun where temperatures are lower than their surroundings, indicative of the Sun's intricate magnetic field. The Sun's plasma, a hot, ionized gas, experiences constant motion due to convective currents, generating magnetic fields through solar dynamo processes \citep{2020LRSP...17....4C}. These areas typically exhibit intense magnetic activity, with concentrated and twisted magnetic fields within sunspots \citep{2009ApJ...702L.133T, 2010ApJ...721..622T}. Electric currents within sunspots are closely tied to the dynamics of these intense magnetic fields.

These electric currents are believed to be driven by the movement of charged particles within the Sun's plasma along the twisted magnetic field lines, contributing significantly to the overall dynamics and behavior of sunspot regions \citep{2014masu.book.....P, 2004psci.book.....A}. Understanding these currents is crucial for comprehending solar activity.

The presence of electric currents in sunspot regions has been associated with flux emergence \citep{1996ApJ...462..547L} and plasma motions in the photosphere \citep{2005A&A...430.1067A}. Such currents deviate from the potential configuration of magnetic fields, playing a significant role in energetic events like solar flares and coronal mass ejections \citep{1993SoPh..148..277A, 1989SoPh..122..215V}.

Theoretical arguments by \citet{1996ApJ...471..485P} suggested that isolated sunspots with twists should have a net zero current, though earlier arguments by \citet{1991ApJ...381..306M, 1995ApJ...451..391M} contradicted this, and \citet{1996ApJ...471..485P} attributing it to unresolved flux tubes. Subsequent studies revealed slightly imbalanced currents in active regions \citep{1996ApJ...462..547L,1998A&A...331..383S, 2000ApJ...532..616W, 2001JGR...10625185F}. However, with the best resolution magnetograms available from space using Hinode/Solar Optical Telescope-Spectro Polarimeter \citep[SOT-SP;][]{2008SoPh..249..167T, 2008SoPh..249..233I}, the absence of net current in isolated sunspots was reported \citep{2009ApJ...706L.114V}. \citet{2012ApJ...761...61G} have reported that the net current is balanced in isolated sunspots and in sunspot groups with intense polarity inversion line exhibits the presence of intense net currents.

Solar flares are believed to be powered by magnetic free energy stored in the current-carrying magnetic fields of active regions \citep{2018SSRv..214...46G}. Studies have found that the presence of non-dominant currents, particularly return currents, is crucial for initiating flares \citep{2011ApJ...740...19R}. Observations have shown strong net currents in highly flare-productive active regions \citep{2018SSRv..214...46G}, with distinct patterns observed between CME-productive and non-productive regions \citep{2017ApJ...846L...6L}. A similar observations in 12 active regions showed that if the ratio of direct to return current is larger than 1.3 it is CME productive \citep{2019MNRAS.486.4936V}.

Understanding the interplay of the magnetic field evolution, in the photosphere, during a flare can provide significant information about the nature of the energy-release process in solar atmosphere and its location.  In this study, we have examined magnetic field evolution associated with the X1.6 class solar flare on October 22, 2014 which occurred in NOAA active region 12192. The objective of our work here is to study in details the change of photospheric magnetic field, and variation of the current in each polarity.  In Section 2, we introduce the observation and data analysis procedure. In Section 3, we present the results followed by discussion and conclusions in Section 4.

\section{Data Analysis}

\begin{figure}[!ht]
\begin{center}
\includegraphics[width=0.7\textwidth]{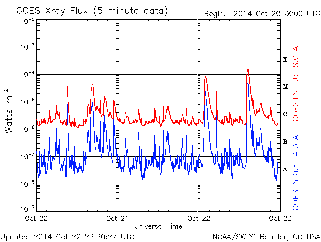} \\
\end{center}
\caption{A plot of GOES soft X-ray flux reveals a series of solar flares occurring between October 20th and 22nd, 2014. These events included a significant X-class flare, as well as multiple C-class and M-class flares.}
\label{fig:1}
\end{figure}

In October 2014, a massive active region NOAA 12192 dominated the Sun's surface. This active region was largest in 25 years, incredibly active, producing a record-breaking series of flares \citep{2019LRSP...16....3T}. During its time facing Earth (October 18th to 29th), NOAA 12192 unleashed a staggering 6 X-class flares and 29 M-class flares. These were accompanied by numerous smaller events. Despite its immense power to produce flares, except some narrow ones \citep{2016ApJ...822L..23P} the region produced no coronal mass ejections (CMEs).

The active region NOAA 12192's immense size wasn't its only distinguishing feature. It possessed a unique magnetic structure, classified as $\beta\gamma\delta$ (Hale classification) and $Fkc$ (McInctosh classification). This complex setup included a leading positive sunspot, two large following negative sunspots, and a mix of magnetic polarities. Notably, the sunspots themselves harbored intricate interactions between opposing magnetic fields.

Prior to rotating into Earth directed view, the region is suspected to have unleashed a powerful CME on October 14th \citep{2015ApJ...801L...6W}. Despite the absence of CMEs during its visible period, several studies have explored why these energetic flares didn't erupt \citep{2015RAA....15.1537J, 2015ApJ...804L..28S, 2015ApJ...809...46C, 2016ApJ...818..168I}. On October 22nd, 2014, NOAA 12192 produced an X1.6 class flare, the strongest type. Figure~\ref{fig:1} shows the GOES soft X-ray (SXR) profiles in the wavelength range 1.0–8.0 (0.4–5.0~\AA). As shown in Figure~\ref{fig:1}, this flare began at 14:02~UT, peaked around 14:28~UT, and ended by 14:50~UT. This spectacular event was observed by numerous spacecraft, including SDO, Hinode, IRIS, and RHESSI. Interestingly, the X1.6 flare was preceded by a weaker C3.2 class flare at 12:00~UT and occurred near the Sun's center (S14, E13). It also did not produce a solar proton event (SPE) or CME.

Studies by \citet{2015ApJ...801L..23T} suggest this flare involved repeated energy releases, hinting at the same magnetic features being repeatedly stressed and reconnected. \citet{2017ApJ...838..134B} investigated the physical properties that contributed to the flare's initiation in the Sun's atmosphere.

To study this event, we used data from two instruments on the Solar Dynamics Observatory \citep[SDO:][]{2012SoPh..275....3P} spacecraft. The Helioseismic and Magnetic Imager \citep[HMI:][]{2012SoPh..275..229S} provides full-disk, high resolution continuum intensity images, magnetograms and dopplergrams. The EUV images/ filtergrams acquired by the Atmospheric Imaging Assembly \citep[AIA:][]{2012SoPh..275...17L} shows the different temperature layers of the solar chromosphere and corona. The photospheric vector magnetograms of the Sun are taken from the ``hmi.sharp$\_$cea$\_$720s data series'' that provides cut out of the active region taken from full-disk vector magnetograms with a 0.5$^{\prime\prime}$ plate scale at a 12 minute cadence \citep{2014SoPh..289.3483H}.

Previous studies indicated that the flare associated magnetic field changes were sufficiently captured with the 12 minutes cadence vector magnetogram data of HMI \citep{2019ApJ...885L..17S} and its noise level is much lower \citep{2017SoPh..292...29L}.  The LOS magnetic field observation from HMI has a pixel resolution 0.5$^{\prime\prime}$ and a cadence of 45~s. The noise level of the LOS field measurement is about 10~G. The HMI vector fields were derived by utilizing the Very Fast Inversion of the Stokes Vector algorithm \citep{2011SoPh..273..267B} based on the Milne–Eddington approximation. On the other hand, AIA, provides full-disk images of the solar corona in a broad UV range with an image scale of 0.6$^{\prime\prime}$ per pixel and a cadence of 12~s covering a wide and nearly continuous coronal temperature in the range of 0.7–20~MK. AIA 17.1~nm and 160~nm images were used to determine timing and location of initial flare brightening.

In the present study, we used de-projected maps in cylindrical equal area coordinates of the automatically identified AR \citep{2014SoPh..289.3549B}. The three prime vector components are B$_{r}$, B$_{p}$, and B$_{t}$. We have downloaded the 12~minute cadence data of vector magnetic field measurements from the JSOC webpage. The vertical magnetic field component (Bz) was utilized for the flux estimation.
\section{Results}

\begin{figure}[!ht]
\begin{center}
\includegraphics[width=0.75\textwidth]{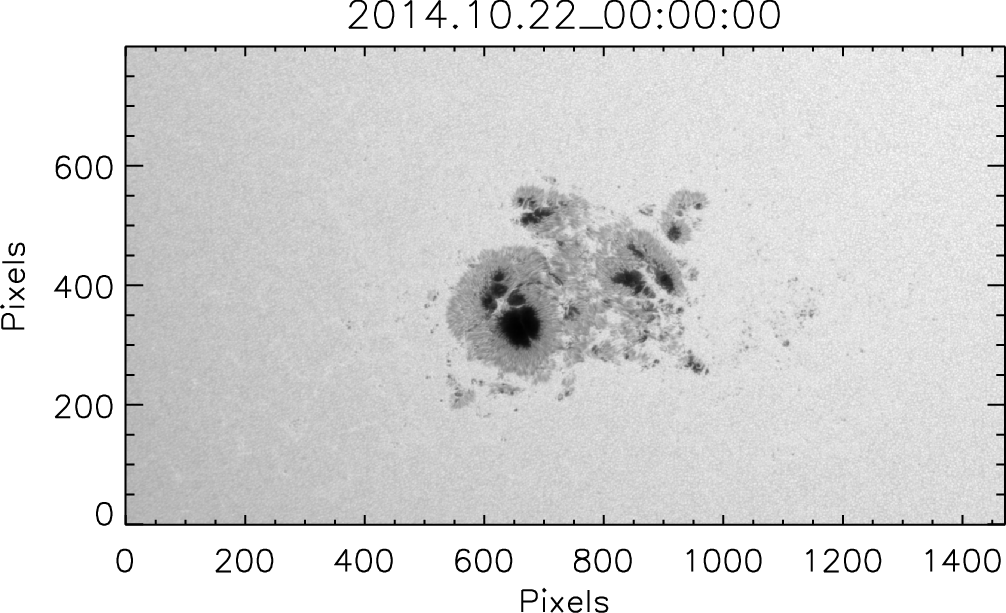} \\
\includegraphics[width=0.75\textwidth]{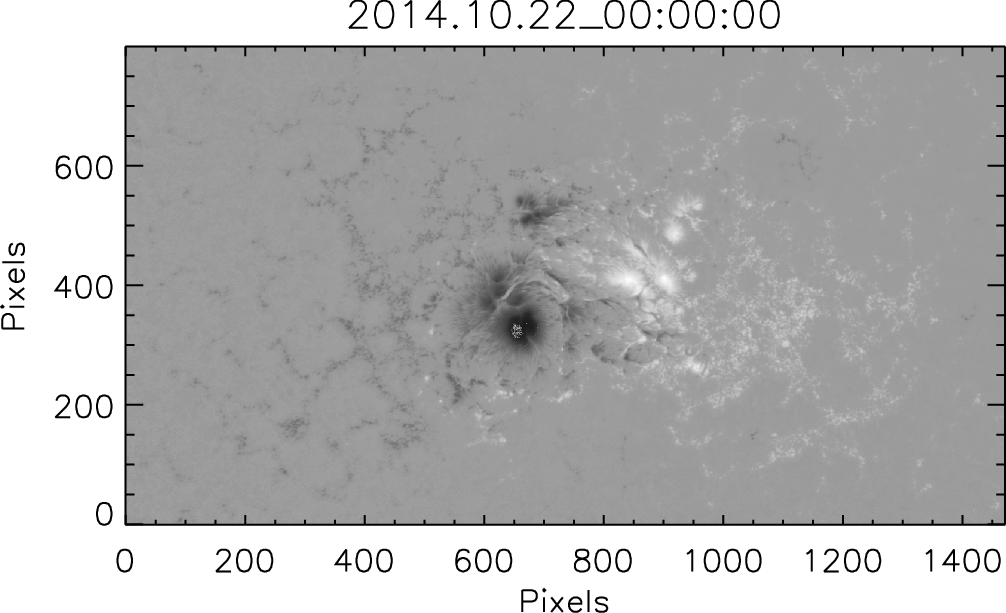} \\
\end{center}
\caption{The top image displays a high-resolution view of Active Region NOAA~12192, captured by the Helioseismic Magnetic Imager Instrument in the Fe~I~6173~\AA~continuum. The bottom image is a corresponding line-of-sight magnetogram of the same region. In the magnetogram, black and white areas indicate regions of south and north magnetic polarity, respectively.}
\label{fig:2}
\end{figure}

\begin{figure}[!ht]
\begin{center}
\includegraphics[width=0.7\textwidth]{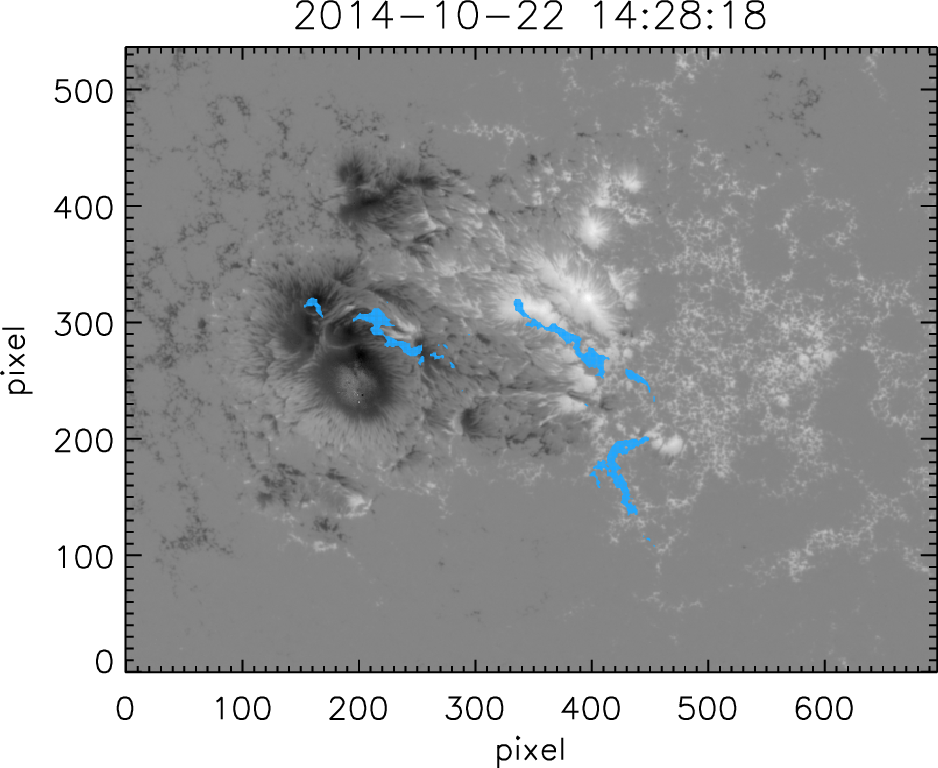} \\
\end{center}
\caption{The image depicts a line-of-sight magnetogram with overlaid bright flare ribbons, highlighted in blue, extracted from 1600~\AA~data. In the magnetogram, black and white areas indicate regions of south and north magnetic polarity, respectively.}
\label{fig:3}
\end{figure}

\begin{figure}[!ht]
\begin{center}
\includegraphics[width=0.6\textwidth]{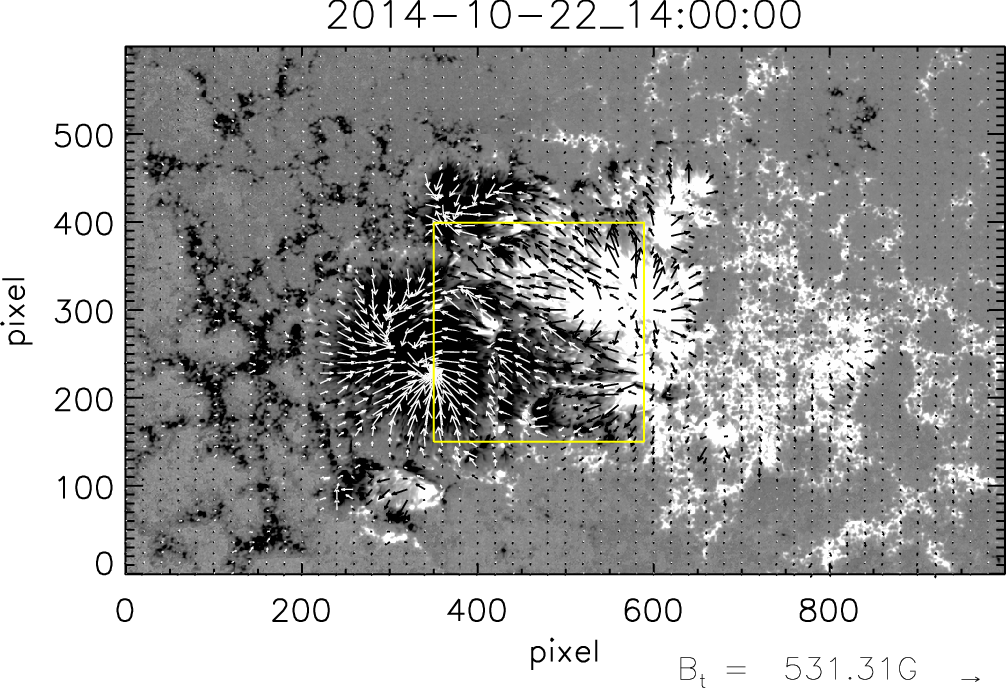}\includegraphics[width=0.4\textwidth]{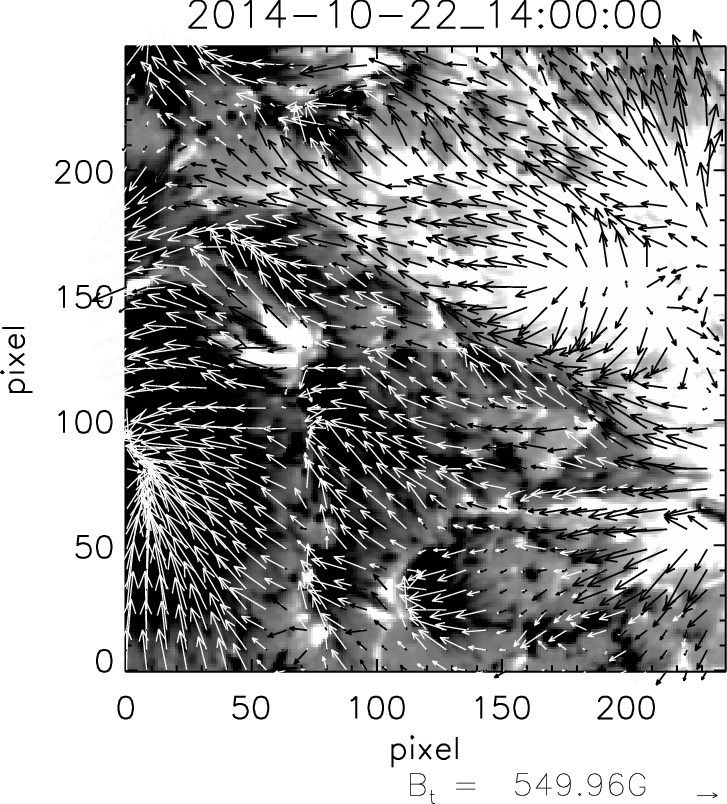} \\
\end{center}
\caption{Left: The horizontal magnetic field vectors are superimposed on the vertical component (Bz) of the magnetic field. The length of each arrow indicates the magnitude of the horizontal field in Gauss. Right: A magnified view of the yellow boxed region shown in the left image.}
\label{fig:4}
\end{figure}

\begin{figure}[!ht]
\begin{center}
\includegraphics[width=0.55\textwidth]{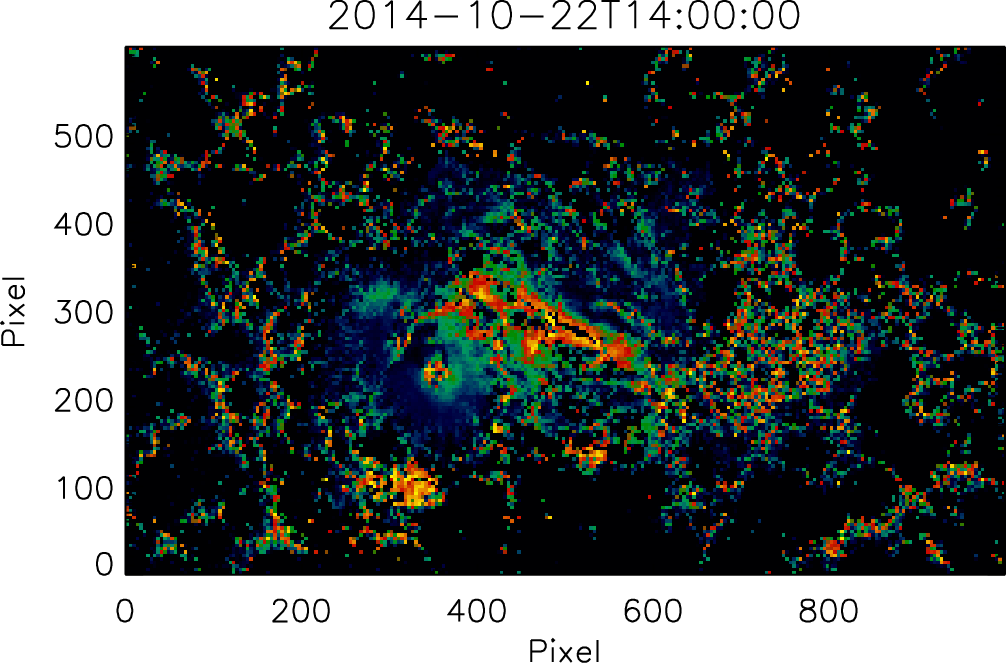}\includegraphics[width=0.45\textwidth]{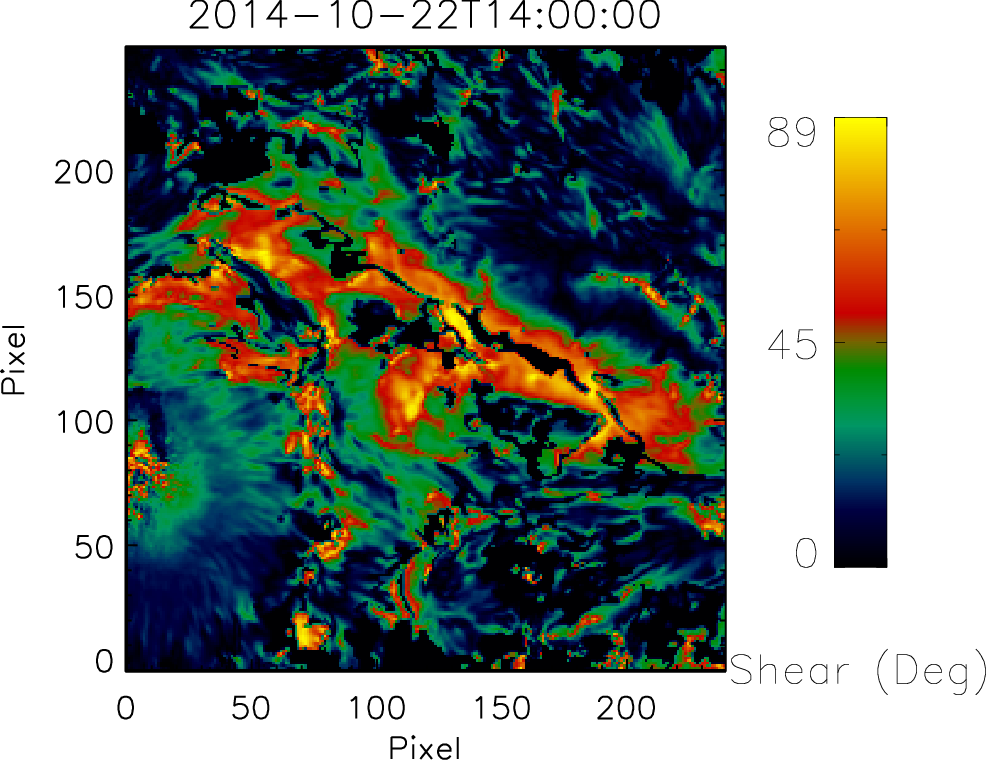} \\
\end{center}
\caption{Left: The shear angle map computed using the observed B$_{x}$ and B$_{y}$ and the computed B$_{xp}$ and B$_{yp}$ using the potential magnetic field configuration.  Right: A magnified view of the shear map of the yellow boxed region shown in the left image of Figure~\ref{fig:4}.}
\label{fig:4a}
\end{figure}

\begin{figure}[!ht]
\begin{center}
\includegraphics[width=0.9\textwidth]{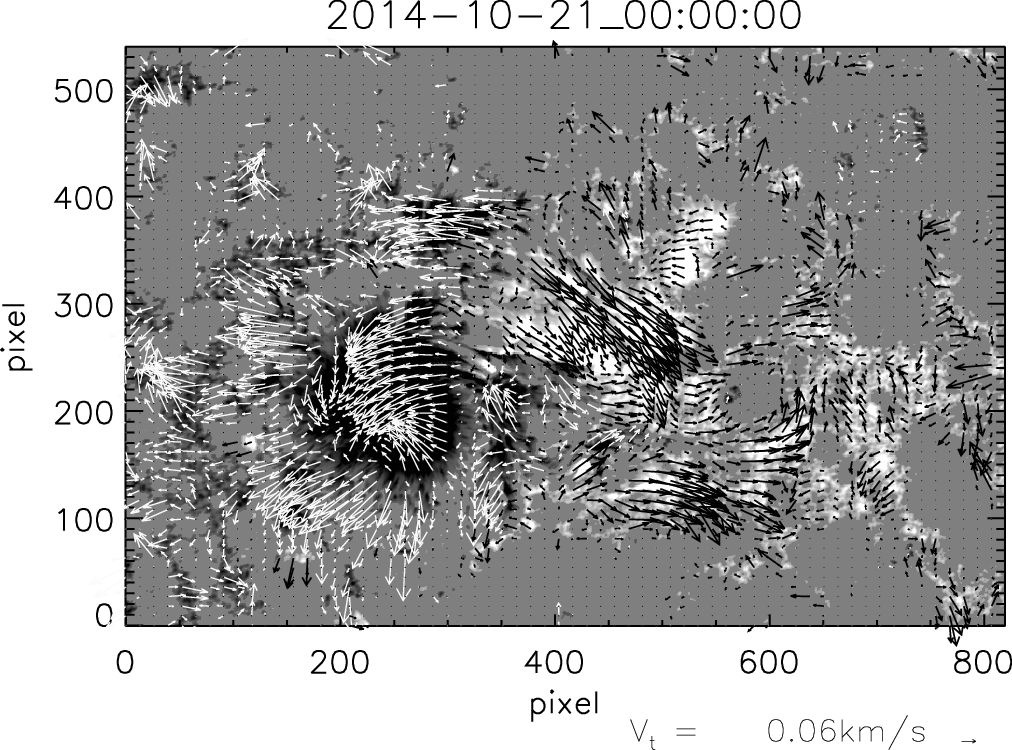} \\
\includegraphics[width=0.9\textwidth]{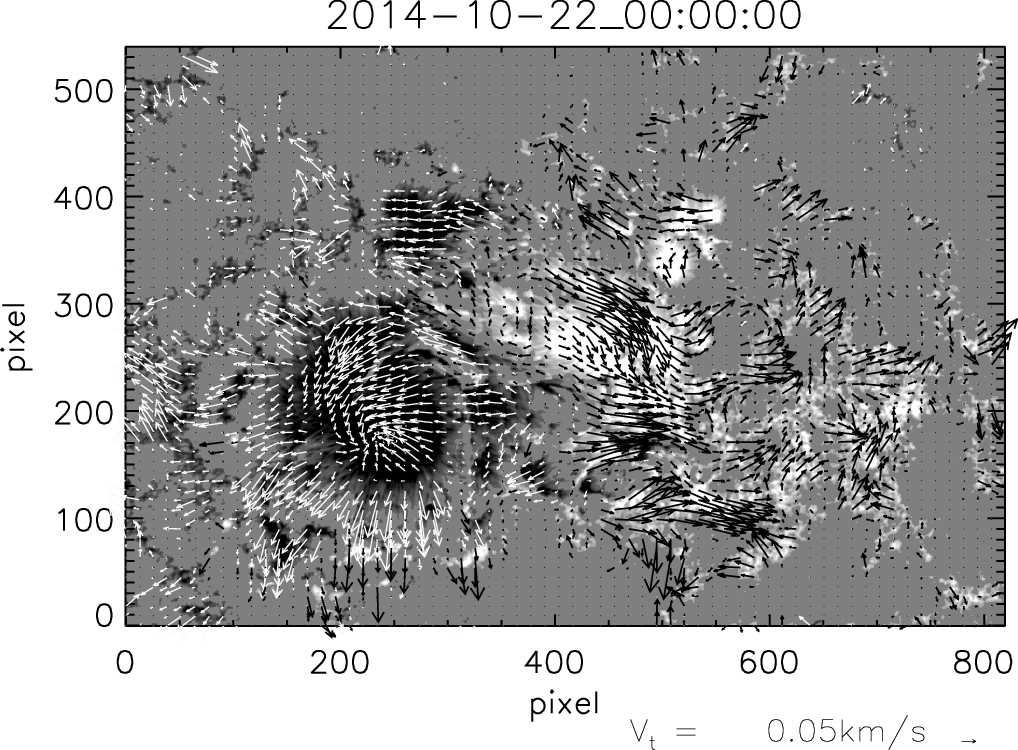} \\
\end{center}
\caption{The horizontal velocity vectors are superimposed on the vertical component (Bz) of the magnetic field. The top and bottom figures depict the same region at different times. The length of arrow in the bottom of the image is indicated in km~s$^{-1}$.}
\label{fig:5}
\end{figure}

\begin{figure}[!ht]
\begin{center}
\includegraphics[width=1.0\textwidth]{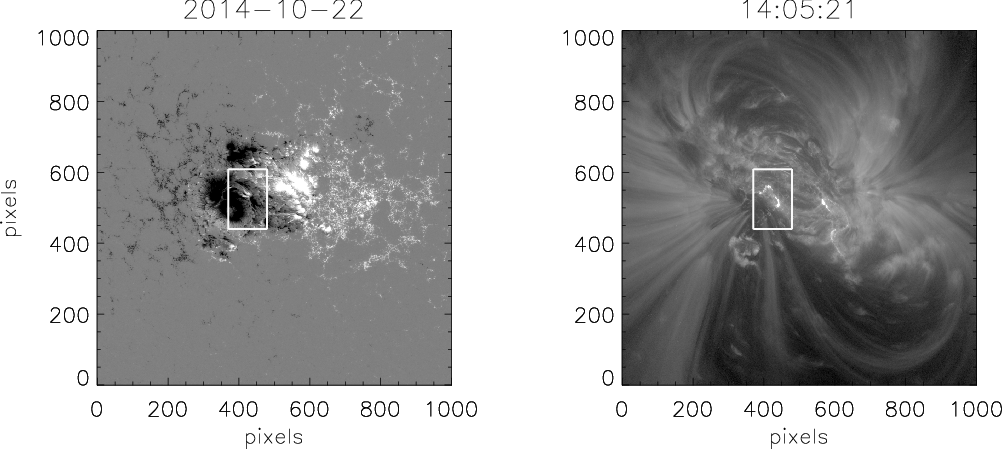} \\
\includegraphics[width=0.45\textwidth]{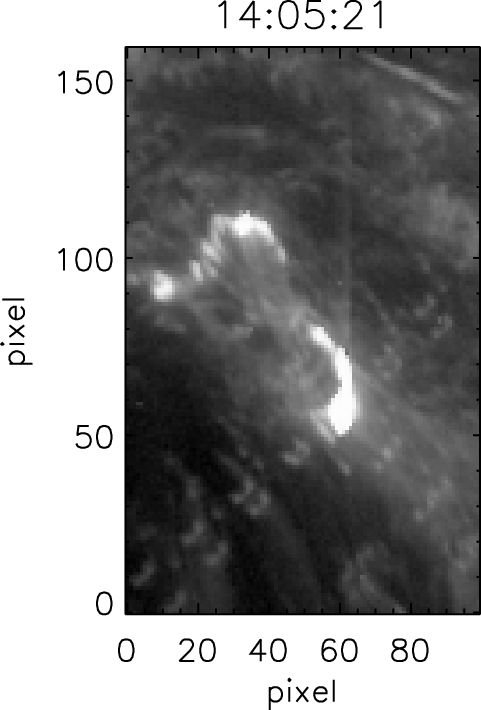} \includegraphics[width=0.45\textwidth]{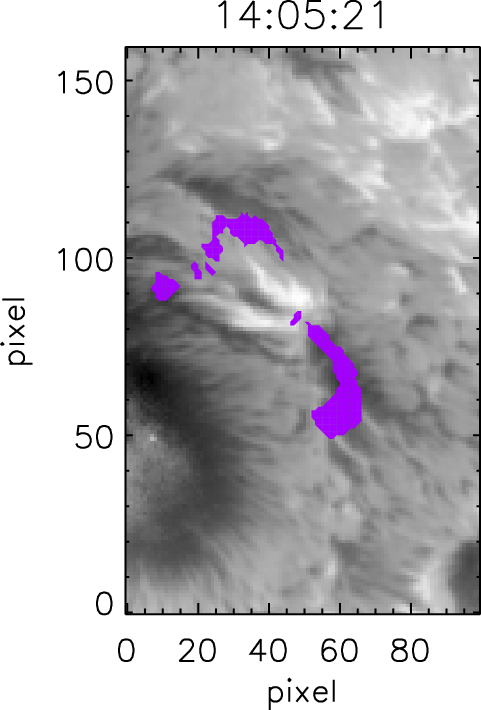}\\
\end{center}
\caption{Top: The image presents the magnetic field configuration within the active region, comprising the line-of-sight magnetic field (left) and the coronal loops (right). Bottom: A brightening observed in the 193~\AA~EUV wavelength (left) during the early stages of the solar flare, and the contour of this brightening superimposed on the magnetogram, emphasizing the emerging flux region where opposite polarity magnetic fields are converging (right).}
\label{fig:6}
\end{figure}

\begin{figure}[!ht]
\begin{center}
\includegraphics[width=0.45\textwidth]{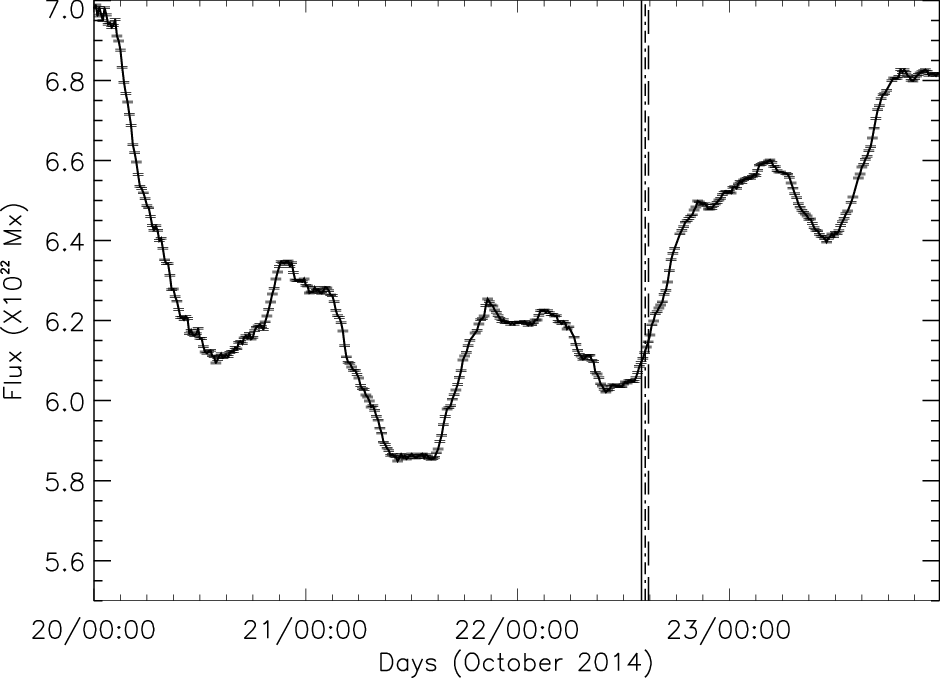}\includegraphics[width=0.45\textwidth]{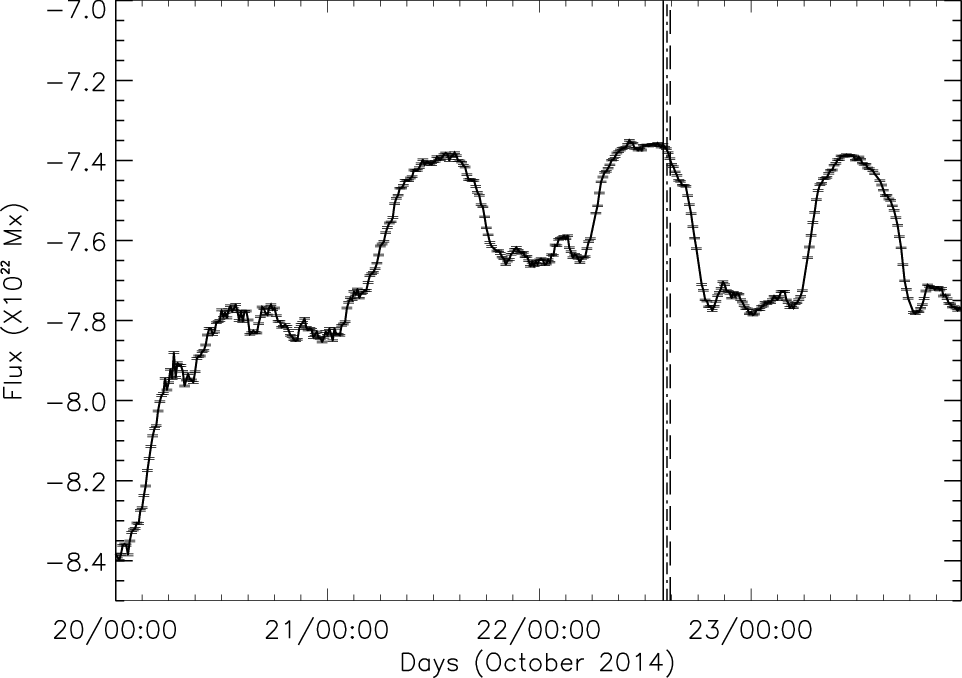} \\
\end{center}
\caption{Magnetic flux was calculated for the positive and negative polarity regions shown in Figure~\ref{fig:6}(top-left). The three vertical lines indicate the onset, peak, and conclusion of the X1.6 class GOES X-ray flare. The small vertical bars on the curve represent the measurement of uncertainty.}
\label{fig:7}
\end{figure}

\begin{figure}[!ht]
\begin{center}
\includegraphics[width=0.8\textwidth]{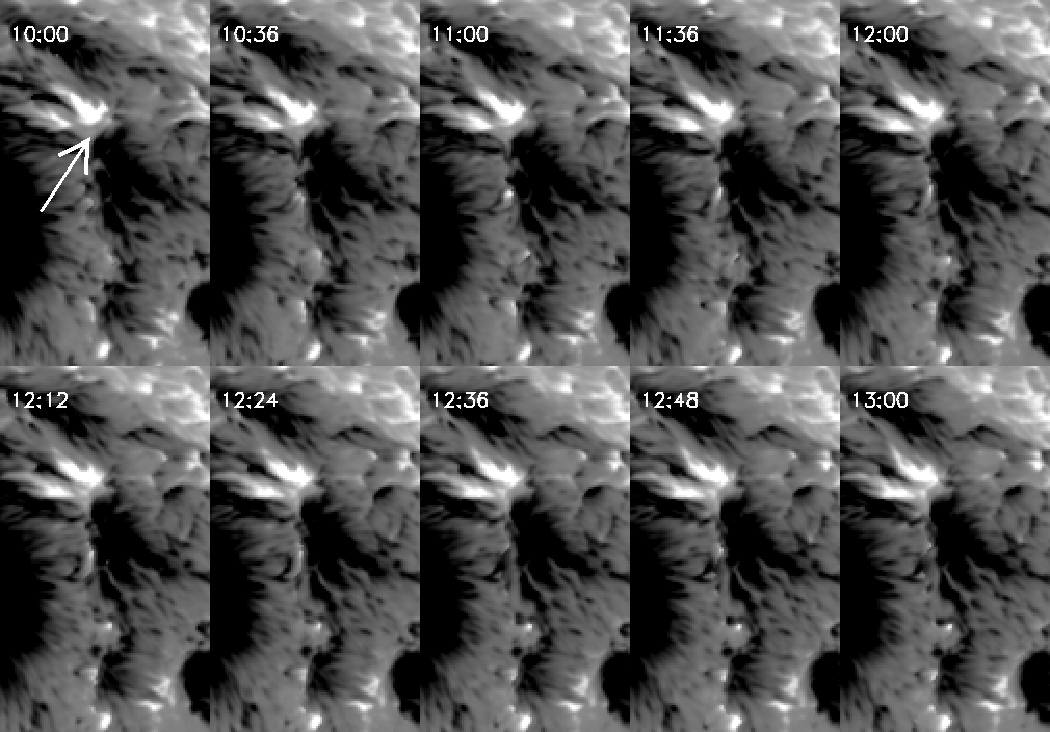}
\end{center}
\caption{The magnetic field evolution is presented for a specific region of interest (indicated by an arrow) where the flare brightening initially occurred on October~22, 2014. The timestamps for each image are displayed at the top.}
\label{fig:8}
\end{figure}

\begin{figure}[!ht]
\begin{center}
\includegraphics[width=0.14\textwidth]{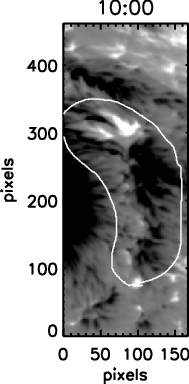}
\includegraphics[width=0.42\textwidth]{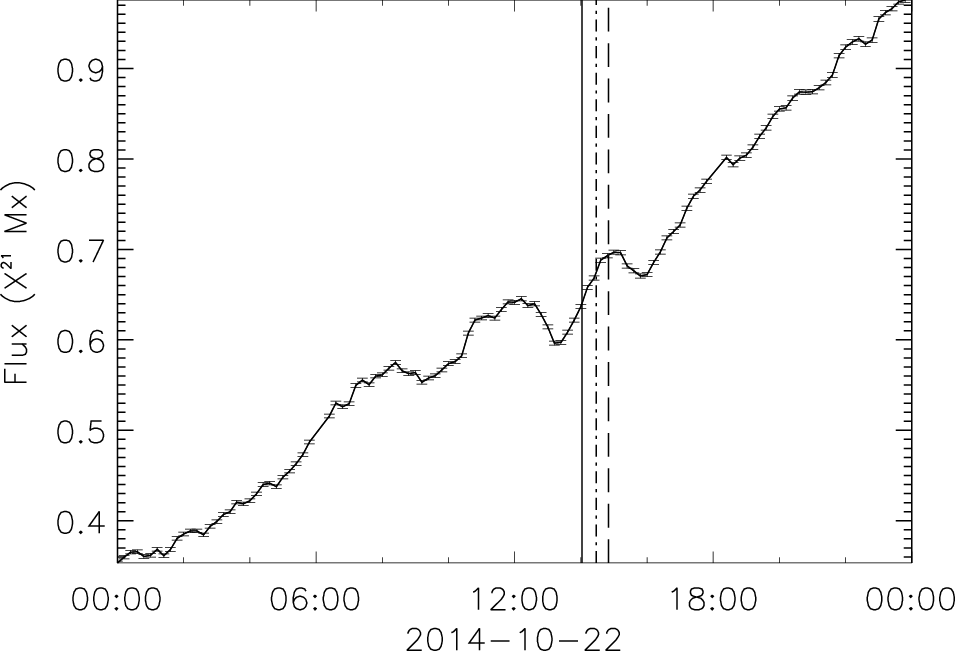}\includegraphics[width=0.42\textwidth]{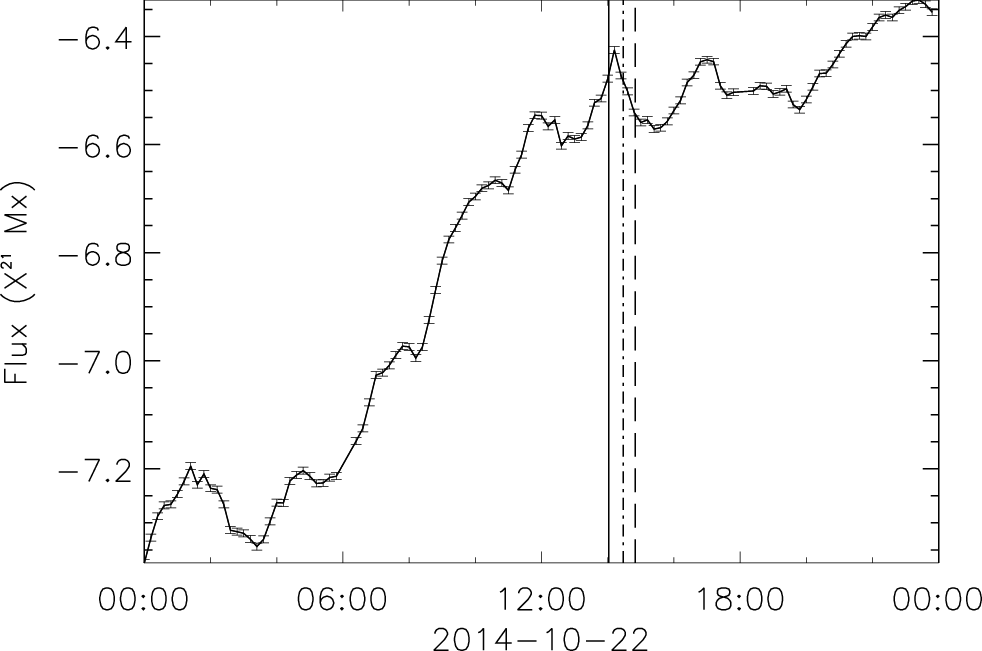} \\
\end{center}
\caption{Left: The region of interest, delineated by contours, was utilized for magnetic flux calculations. The middle plot illustrates the temporal evolution of positive flux, while the right plot depicts the negative flux. The 3 vertical bars indicates the beginning, peak and end time of the flare. While, the small vertical bars on the curve represent the measurement of uncertainty.}
\label{fig:9}
\end{figure}

\begin{figure}[!ht]
\begin{center}
\includegraphics[width=0.9\textwidth]{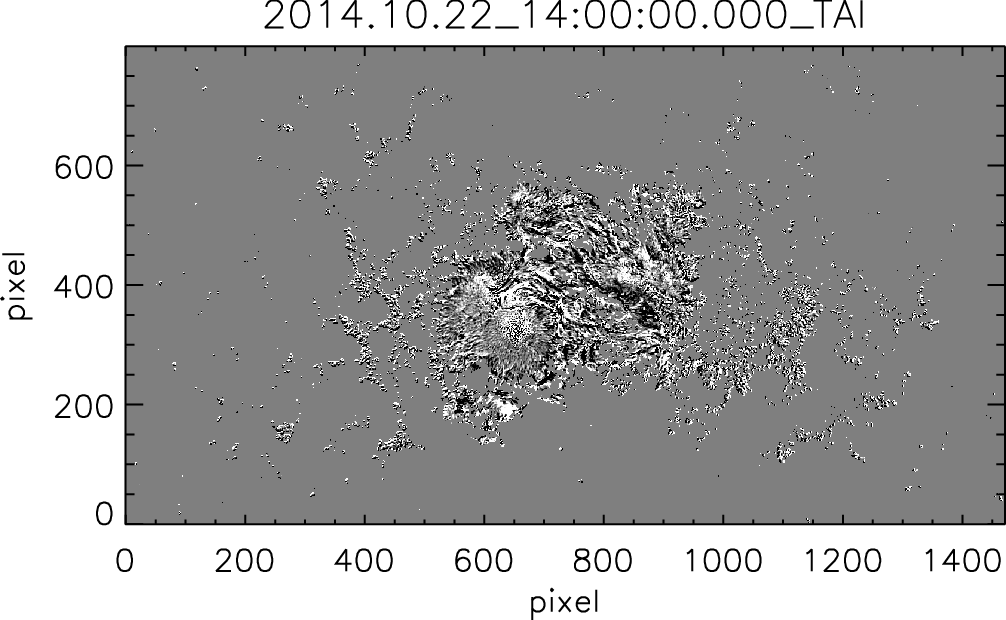} \\
\end{center}
\caption{The vertical current density (J$_{z}$) map of NOAA~AR~12192 was calculated using the horizontal components of the vector magnetic field. Black and white areas indicate regions of opposite J$_{z}$ signs.}
\label{fig:10}
\end{figure}

\begin{figure}[!ht]
\begin{center}
\includegraphics[width=0.9\textwidth]{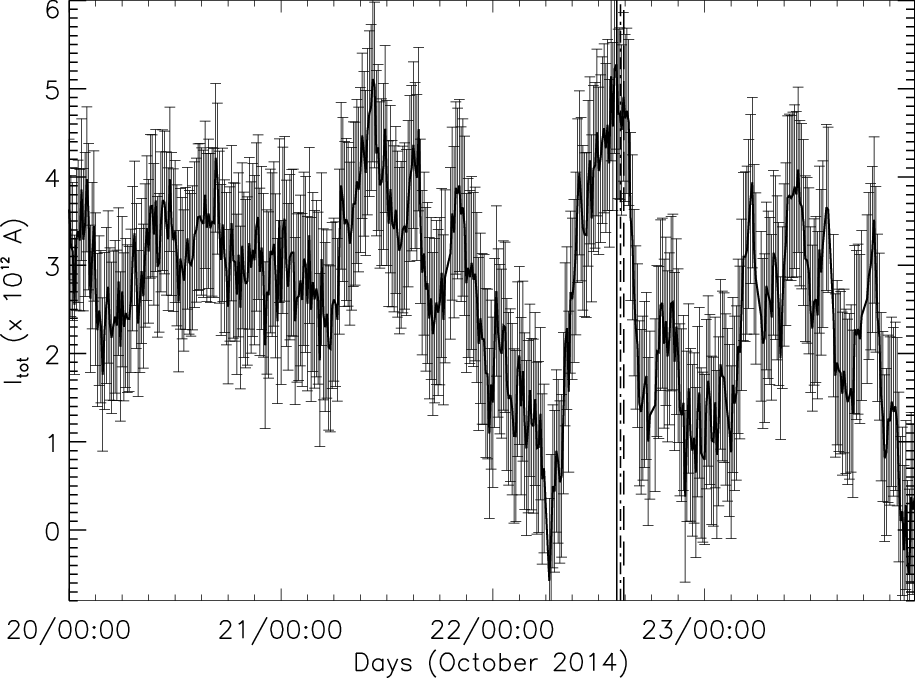} \\
\end{center}
\caption{The plot depicts the temporal evolution of the total current over a four-day period. The three vertical lines indicate the onset, peak, and conclusion of the X1.6 class flare. The vertical bars on the curve represent the measurement of uncertainty.}
\label{fig:11}
\end{figure}

\begin{figure}[!ht]
\begin{center}
\includegraphics[width=0.9\textwidth]{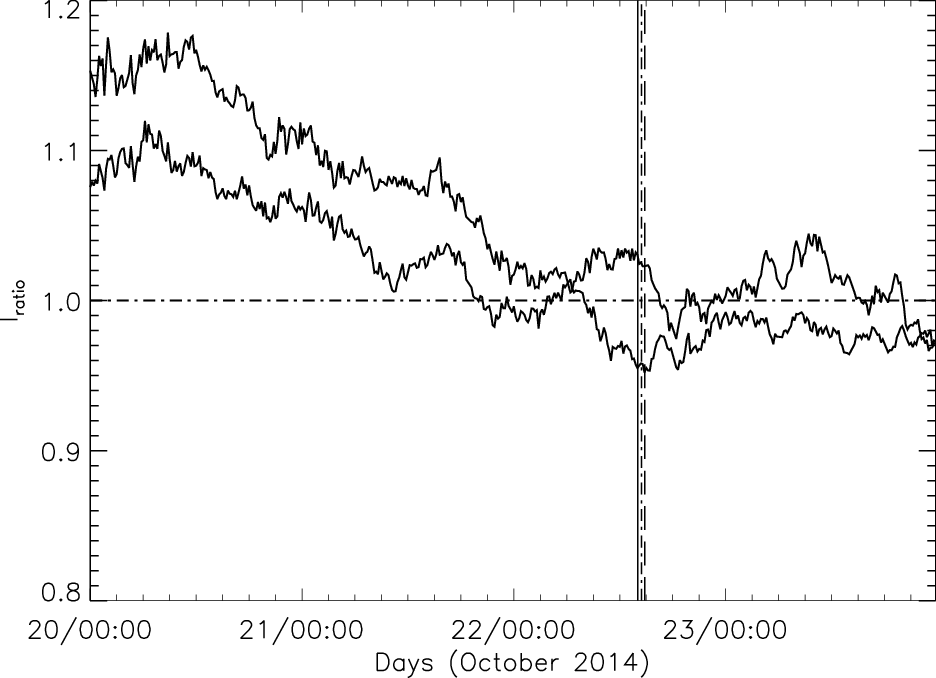} \\
\end{center}
\caption{The plot illustrates the temporal evolution of the ratio of DC to RC current within each magnetic polarity. The three vertical lines indicate the onset, peak, and conclusion of the X1.6 class flare. The horizontal line at a value of 1 represents the point where DC and RC values are equal.}
\label{fig:12}
\end{figure}

\begin{figure}[!ht]
\begin{center}
\includegraphics[width=0.9\textwidth]{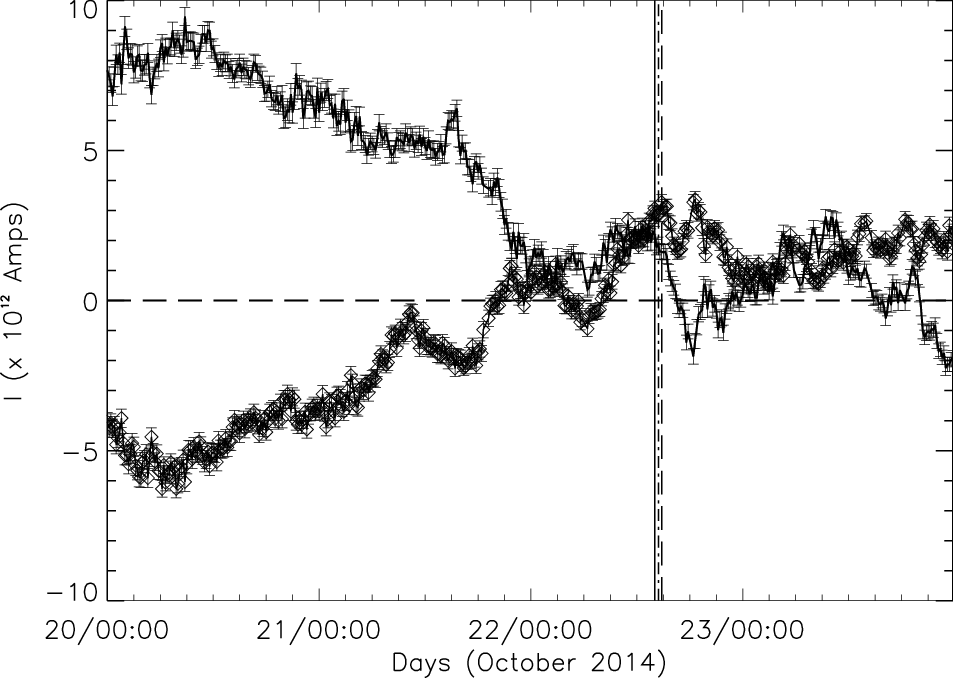} \\
\end{center}
\caption{The plot depicts the temporal evolution of the total current within the north polarity region (solid line) and south polarity region (diamond symbol) over a four-day period, encompassing the X1.6 class flare. The three vertical lines align with those in the previous plots, indicating the onset, peak, and end of the flare. The vertical bars on the curve represent the measurement of uncertainty.}
\label{fig:13}
\end{figure}

A large sunspot group AR NOAA 12192, appeared on the Sun's eastern limb in the southern hemisphere at a latitude of 12$^{\circ}$  on October 16th, 2014. By the time it was first observed, it had already developed into a substantial region classified as beta-gamma class. The sunspot group crossed the western edge of the Sun on October 29th, 2014. At its peak, it covered an impressive 2400 millionths of the Sun's visible hemisphere.

Figure~\ref{fig:2}(top) shows AR NOAA 12192 in detail, observed in the continuum of the Fe~I~6173~\AA~line. The corresponding magnetogram is displayed in the bottom  of Figure~\ref{fig:2}. As seen in both images, AR NOAA 12192 consists of a single dominant sunspot with several smaller companions. The smaller spots exhibit a northern magnetic polarity, while the larger central spot has a southern magnetic polarity. A closer look at the magnetogram indicates that this large, negative polarity sunspot group has a complex structure. It contains multiple dark cores (umbrae) separated by light bridges and is surrounded by penumbra that shares the same magnetic polarity.

The time series of HMI magnetograms header provide the averaged and integrated quantities of several magnetic parameters \citep{2020A&A...639A..44H}. Since integrated and mean quantities provides the values across the entire active region, they can obscure variations within the sunspot group. To capture these local variations and gain a deeper insight into the magnetic field, we analyzed specific parameters at a smaller spatial scale. The detailed results are presented in the following section.

 Figure~\ref{fig:3} displays the active region with the flare ribbons superimposed on a line-of-sight magnetogram. Notably, the ribbons are not parallel, but rather appear on opposite sides – one in the southern polarity region and the other in the northern polarity region, both close to the polarity inversion line. As observed in the AIA/SDO filtergram at 1600~\AA, the initial brightening of the flare ribbons began at 14:05:06 UT. This timing aligns with what was observed in the AIA/SDO images at 1700~\AA. However, the brightening in the corona, captured at a shorter wavelength of 193~\AA, started a little earlier at 14:04:56 UT.

Figure~\ref{fig:4}(left) depicts the horizontal magnetic field vectors overlaid on the vertical component (Bz) of the magnetic field. The arrow lengths represent the magnetic field strength in Gauss. The region between the opposing polarities exhibits strong shear, with the field vectors nearly parallel to the neutral line. This can be seen in the zoomed in view of the boxed region (yellow color) near the neutral line in Figure~\ref{fig:4}(right).

To demonstrate the presence of magnetic shear near the polarity inversion line, we computed the angle between the observed vector field and the potential magnetic field \citep{1989SSRv...51...11S}. Figure~\ref{fig:4a}(left) presents the shear map for the region depicted in Figure~\ref{fig:4}(left), while Figure~\ref{fig:4a}(right) shows the shear map for the region in Figure~\ref{fig:4}(right). A clear enhancement in the shear angle is observed near the polarity inversion line on both sides of the PIL. The mean shear angle for the smaller region is 23$^{\circ}$. This shear angle is consistent with the value obtained by \citet{2024ApJ...961..148L}. However, this value is influenced by numerous pixels with low shear angles (black and blue colored regions). By focusing on the areas very close to the polarity inversion line, we find a significantly higher mean shear angle of 65$^{\circ}$, indicating a highly sheared region near the PIL.

This active region displayed continuous growth from its emergence on the Sun's Eastern limb. While the negative polarity region was already well-developed, emergence of magnetic flux continued in both polarities. To study the motion of magnetic features in this active region, we applied local correlation tracking \citep[LCT:][]{1988ApJ...333..427N} technique to continuum images taken 12 minutes apart. A 15-arcsec pixel apodizing window function was used. The resulting horizontal velocity vectors are overlaid on the Bz component of the magnetic field. Figure~\ref{fig:5} shows the flow field for October 21 and 22. In addition to the typical outward flow seen in sunspot penumbra, we observed an inward flow moving in a counterclockwise direction in the center of sunspot. The positive polarity region exhibited southwestward flow. We suggest that this complex evolution of magnetic fields could have caused this X-class flare.

The two flare ribbons originated in regions of oppositely oriented magnetic fields within the sunspot group. Before the flare erupts, a set of hot loops connects these two regions. Figure~\ref{fig:6}(top-right) illustrates this loop configuration in the corona, clearly showing its connection from the sunspot's north magnetic pole to the south pole. The bottom images in Figure~\ref{fig:6} provide a closer look at the neutral line. A contour of the pre-flare brightening region is overlaid on the magnified view of the magnetogram in Figure~\ref{fig:6}. This region corresponds to the boxed region in the top images of the same figure. Pre-flare brightening was observed near the neutral line and in the negative polarity region at 14:05:21 UT in the 193~\AA~wavelength (sensitive to hot plasma at about 1.5~MK \citep{2012SoPh..275...17L}). This spatial association strongly suggests that the flare originated close to the neutral line of the mustache-type positive polarity region (Figure~\ref{fig:9} contoured region).

\subsection{Magnetic Flux Evolution}
Magnetic flux is a measure of the total amount of magnetic field passing through a surface.
We calculated the magnetic flux separately for the positive and negative polarity regions in AR NOAA 12192, considering only pixels with a z-component of the magnetic field strength exceeding 100~G (Figure~\ref{fig:7}). The SDO/HMI measures the vertical magnetic field strength with an uncertainty of about 100G \citep{2014SoPh..289.3483H}. This filtering helps exclude noisy data and pixels outside the active region which may not contribute to the overall evolution of the active region. Especially since we are measuring the field in a strong magnetic-field AR, this cut off of 100~G is deemed suitable. During the flare, the active region was close to the disk center (S14E13) and we used B$_{z}$ component of the vector field in the computation of flux. Hence, the projection effect do not affect our flux calculation \citep[see also:][]{2016ApJ...833L..31F}.

As observed in the continuum image, the negative polarity region was larger than the positive one. Consequently, the total magnetic flux in the negative polarity was about 1.2 times greater. Interestingly, the flux in both polarities exhibited a similar pattern throughout the observations. During the X1.6 class flare itself, the magnetic flux values were approximately 6.1$\times$10$^{22}$~Mx for the positive polarity and 7.4~$\times$~10$^{22}$~Mx for the negative polarity. However, a crucial observation is that the flux actually increased during the flare. 

A small region of positive and negative polarity, marked by an arrow in Figure~\ref{fig:8}, exhibits continuous but slow flux emergence, particularly evident in the growing positive area. A clearer view of flux emergence is provided in Figure~\ref{fig:9}, where the region is isolated using DEFROI.PRO IDL function for flux calculations within each polarity and is shown in the left-side image with contour. The middle plot shows positive flux, and the right plot shows negative flux. Both polarities display a temporal increase in flux suggesting the continuous flux emergence. This result is similar to the observations of \citet{2004ApJ...605..931W}.

Intriguingly, the southern polarity's flux begins to rise just after the flare initiation and continues even after the flare subsides. Conversely, the northern polarity's flux initially decreases after the flare but then exhibits a delayed rise and fall pattern approximately 30 minutes later.

\subsection{Electric Current Density}
Understanding electric currents within active regions during flares is crucial because they're closely linked to the presence of a `non-potential' magnetic field configuration, which can store significant energy. Unfortunately, directly measuring these currents in the Sun's atmosphere is impossible. The only approach available relies on the vector magnetic field. By applying Ampere's Law, the curl of this magnetic field allows us to indirectly calculate the electric current density in the photosphere. In essence, the hot plasma flow within the active region distorts the magnetic field. This distortion, as described by Ampere's Law, generates an electric current. To compute the vertical current density (J$_{z}$), we utilize the horizontal components of the magnetic field (B$_{x}$ and B$_{y}$) as shown in the following equation

\begin{equation}
\frac{1}{\mu_{0}}(\frac{\partial{B}_{y}}{\partial{x}}-\frac{\partial{B}_{x}}{\partial{y}}).
\end{equation}

Where $\mu_{0}=4\pi\times10^{-3}~G~mA^{-1}$
is the permeability in free space and the unit of J$_{z}$ is Am$^{-2}$. The vertical component of the current density J$_{z}$ is composed of two terms, the shear and the twist \citep{2001ApJ...557L..71Z}. This vertical current density (J$_{z}$) measure accounts for horizontal field gradients and is calculated with the observations of the vector field.

Figure~\ref{fig:10} displays a map of the electric current density within the active region, calculated using vector magnetogram obtained at 14:00~UT. This figure reveals the distribution of the vertical current density (J$_{z}$) across different locations in the active region. We can see both negative and positive J$_{z}$ values distributed throughout the region. The J$_{z}$ distribution exhibits a `salt and pepper' pattern in the darker, central umbrae of sunspots and the quieter Sun regions. In the penumbral regions, the pattern appears more fibrilar. However, the data quality in these penumbral regions is not as good as what has been observed with the Hinode vector magnetograph \citep{2009ApJ...697L.103S, 2009ApJ...706L.114V, 2011ApJ...740...19R}. Notably, certain areas within the active region are dominated by negative current density, while others are dominated by positive current density.

Figure~\ref{fig:11} illustrates the total vertical current flowing through the active region over time. This value is obtained by integrating the current density (J$_{z}$) across the entire sunspot area. To minimize the influence of noise and weak magnetic fields, we only considered pixels where the vertical field strength exceeded 100 Gauss and the horizontal field strength exceeded 250 Gauss during this integration. This is to select only the strong field pixels whose total field strength is larger than 250~G. This is 2 times larger than the noise in total magnetic field strength of HMI vector magnetic field measurements \citep{2014SoPh..289.3483H, 2017ApJ...846L...6L}.

As seen in Figure~\ref{fig:11}, the total current remained around 3~trillion Amps for the first two days. However, it began to rise in the middle of October 21st. Interestingly, the current direction (positive or negative) switched during the early hours of October 22nd. It then became positive and exhibited a significant increase in magnitude over the next 9 hours. Notably, the X1.6 class flare erupted when the total current reached 5~Terra Amps. Following the flare, the net current started to decrease. The vertical bar represents the error in computing the net current \citep{2011ApJ...740...19R}. The electric current in a magnetic system indicates the deviation of the magnetic ﬁeld from its potential conﬁguration, referring to the accumulation of free energy. From the figure the  decrease of current after the X1.6 flare could be related to the field approaching potential configuration.

Figure~\ref{fig:12} depicts the temporal evolution of the dominant-to-non-dominant current ratio. In the plot it is seen that the ratio of dominant to non dominant current is only a little larger than 1 before 22nd, hence could not have seen any flare producing CMEs \citep{2015ApJ...808L..24C, 2016ApJ...822L..23P}. Then it became close to 1 after 22nd (Figure~\ref{fig:12}). In the same sense the flare could be confined flare when ratio is close to 1 as has been seen in \citet{2017ApJ...846L...6L}.

Figure~\ref{fig:13} dives deeper by separating the total current calculations for the northern and southern magnetic polarity regions. This plot tracks the current in each region over a four-day period, from October 20th to 23rd, 2014. For most of October 21st, the total current in the N-polarity region flowed in a positive direction, while the S-polarity region carried a negative current. On October 20th, the N-polarity region started with a net current of around $\sim$7$\times$10$^{12}$~Amps, while the S-polarity region had a net current of roughly $\sim$4$\times$10$^{12}$~Amps. These values peaked later that day, reaching approximately 9$\times$10$^{12}$~Amps in the N-polarity region and 6.1$\times$10$^{12}$~Amps in the S-polarity region. The plot reveals that both currents with opposite directions began to decrease from October 21st onwards.

Interestingly, the current in the S-polarity region flipped its sign from negative to positive around 00:00~UT of October 22nd and remained positive until 15.00~UT. Notably, the X1.6 class flare erupted when the total current in both polarities shared the same positive sign.

Following the flare's end, the total current in the N-polarity region reversed its direction again, transitioning from negative to positive. This shift coincided with another flare, an M1.1 class flare observed on October 23rd, 2014. Our analysis suggests that the X1.6 class flare might be linked to a specific condition: when the currents in opposite magnetic polarities not only have the same sign but also reach their peak values simultaneously. This potentially creates an environment conducive to magnetic reconnection, a key process in solar flares.

\section{Summary and Discussions} 
We investigated active region NOAA 12192 during the powerful X1.6 flare it produced on October 22, 2014. Using data from the SDO spacecraft, we analyzed photospheric vector magnetograms to understand this event. We focused on identifying magnetic parameters in these magnetograms that correlate with the morphological changes in NOAA 12192 during the flare. Here is the summary of our findings.

1] Active region NOAA 12192 exhibited dynamic changes in magnetic field during its time on the solar disk. The observed decrease and subsequent increase in magnetic flux suggests ongoing cycles of magnetic flux cancellation and emergence. The horizontal magnetic field is aligned along the polarity inversion line (PIL). Additionally, the negative polarity spot exhibits a converging anticlockwise flow, while the positive polarity region shows southward movement.

2] During the X1.6 flare, we observed a significant increase in magnetic flux in both the positive and negative polarity regions of active region 12192. The positive polarity reached a value of approximately 6.1$\times$10$^{22}$~Mx, while the negative polarity reached approximately 7.4$\times$10$^{22}$ Mx. This suggests an emergence of magnetic flux in both polarities, potentially contributing to the energy released during the flare.

3] The SDO vector magnetograms, analyzed at 14:00 UT, revealed a map of both positive and negative vertical current density (J$_{z}$) distributed throughout the active region. This pattern resembled a ``salt and pepper'' distribution in the umbra and quieter areas, transitioning to a ``fibril'' pattern in the penumbra. This finding aligns with observations of J$_{z}$ from high-resolution Hinode magnetograms \citep{2009ApJ...702L.133T, 2011ApJ...740...19R, 2012ApJ...744...65T}. Notably, some areas within the active region displayed dominance of negative current density, while others exhibited dominance of positive current density.

4] Our analysis of the total current revealed a significant increase in its magnitude after October 21st, accompanied by a sign change. Notably, the total current reached 5 Terra Amps during the X1.6 flare, then began to decrease afterwards. Interestingly, the plot of net current from October 22th to end of 23rd indicated that the flare occurred when the total current in both polarities had the same sign. During the flare onset, the net current in the positive polarity region was around 2$\times$~10$^{12}$~Amps, while the negative polarity held a value of approximately 3$\times$10$^{12}$~Amps. It's worth noting that the total current in the northern polarity region flipped its sign after the flare.

Prior to the flare, analysis of the active region using vector magnetograms revealed sunspots of both north (N) and south (S) polarity. The S-polarity sunspot was larger and well-defined, while the N-polarity region was smaller and appeared to be emerging. Notably, the transverse magnetic field lines exhibited significant shearing near the boundaries where the polarities reversed (polarity inversion lines, PILs). We have observed increase in magnetic flux within both polarities before the flare. This increase in flux is likely created an imbalance, setting the stage for magnetic reconnection, a key process in solar flares. Additionally, the S-polarity sunspot held a greater amount of magnetic flux compared to the N-polarity region.

\citet{2003A&A...406..337T} reported a significant decrease in the net magnetic flux of active regions 2-3 days before major flares, accompanied by a reduction in flux imbalance. Similarly, \citet{2013AdSpR..52.1561C} observed a decrease in flux imbalance within NOAA 11283 prior to a major flare in September 2011. These studies suggest a potential link between a significant decrease in flux imbalance and the destabilization of large-scale magnetic fields in active regions, which may contribute to the occurrence of powerful flares.

Several research articles highlight the importance of the ratio of direct (DC) to return current (RC) in predicting CMEs \citep{2017ApJ...846L...6L, 2019MNRAS.486.4936V, 2020ApJ...893..123A, 2024ApJ...961..148L}. This ratio serves as a proxy for the non-neutralization of current in the Sun's active region. A ratio greater than 1 indicates a region with significant non-neutralized current, suggesting a higher likelihood of an eruptive flare and a potential CME. Conversely, a ratio less than or close to 1 implies a well-neutralized current, making the flare less likely to erupt. This concept stems from the observation that a DC/RC ratio exceeding 1 is often accompanied by strong shear flows near the neutral line in the magnetic field \citep{2015ApJ...810...17D}. These shear flows, produces shear angles  exceeding 60 degrees, can destabilize the magnetic configuration, leading to an eruption. For instance, in Active Region (AR) 12192, our analysis \citep[see also][]{2017ApJ...846L...6L, 2019MNRAS.486.4936V} revealed a ratio slightly above 1 (but below 1.2) before October 21, 2014. This period witnessed a few flares, but no CMEs. Interestingly, the ratio approached 1 after October 22nd, coinciding with the occurrence of an X1.1 class flare and other events. This observation aligns with the notion that a well-neutralized current (ratio close to 1) is less prone to eruptive activity.

\citet{2015ApJ...804L..28S} further support this concept by characterizing AR 12192 as a large region with weak non-potentiality and a strong overlying magnetic field. With the 3D MHD numerical simulations \citet{2015ApJ...810...17D} also demonstrate that net currents develop within the corona when magnetic twisting and shearing motions occur near the polarity inversion line. These currents can significantly influence the stability of the magnetic field and can cause flares. In essence, the DC/RC ratio provides a valuable tool for understanding the eruptive potential of active regions. A high ratio indicates a higher risk of CMEs, while a low ratio suggests a more stable configuration with a lower CME risk.

\citet{2015ApJ...808L..24C, 2019ApJ...885...89V}, have explored a parameter called the coronal horizontal background magnetic field  decay index ($n$) to understand solar eruptions. This index reflects the weakening of the Sun's magnetic field with height in the corona, measured over a specific active region and time frame (7 days in Chen et al.'s study, at an altitude range of 45-105 Mm). Their observations for Active Region (AR) 12192 revealed an $n$ value between 1.2 and 1.5 from October 20th to 26th, 2014. A value greater than 1.5 is often linked to eruptive events, like coronal mass ejections (CMEs), while a value below 1.5 indicates a more stable configuration, less likely to erupt.  In AR 12192, the $n$ value is above 1.2 and there were no CMEs observed. \citet{2015ApJ...808L..24C} proposed that it could be due to the strong overlying background magnetic field. Imagine the magnetic field lines as loops. A strong overlying field acts like a powerful lid, suppressing the eruption of these loops even when the internal magnetic forces might be pushing for an outburst. This confinement by the overlying field explains the lack of CMEs in AR 12192.

\citet{2016ApJ...828...62J} used data-driven numerical magnetohydrodynamic (MHD) modeling to study a X3.1 class solar flare in AR 12192. The model successfully reproduced the macroscopic magnetic process of the flare, where a large-scale coronal current sheet was created and triggered reconnection, resulting in an impulsive release of magnetic energy in AR 12192. The simulated results matched well with the observations, including the location of chromospheric flare ribbons and the morphology of reconnecting field lines with no CMEs.

\citet{2011ApJ...740...19R} studied current evolution in solar active region 10930. They found that initially, net current in each polarity increased with magnetic flux, then decreased. This decrease coincided with the rise of a current opposite the net current. The dominant current (same sign as net current) peaked much before major flares. They proposed that changes in the non-dominant current are crucial for flare initiation, possibly due to Lorentz force driving magnetic reconnection.

The electric current in the active region can produce magnetic field in different directions leading to imbalance of magnetic flux. Our analysis revealed a decreasing trend in the total current of the active region starting from late October 21st. Interestingly, this decrease was followed by a sudden rise just before the X1.6 flare. During the flare itself, the total current reached a magnitude of approximately 5$\times$10$^{12}$~Amps. This pre-flare rise in net current suggests the emergence of new magnetic flux within the active region. Furthermore, we observed a decrease in the net current of both opposing polarities for two days preceding the flare. This reduction can likely be attributed to an increase in the opposing current within each respective polarity region. This dynamic behavior of the net current in opposite polarities might be linked to the evolving deformations of the magnetic field in the flaring region.

Importantly, \cite{2014masu.book.....P} suggests that such dynamic magnetic field evolution is crucial for triggering flares through reconnection. Based on our findings, we propose that magnetic reconnection in the flaring site occurred when the total currents in both polarities reached their peak values and became aligned in the same direction. This potential alignment of oppositely directed magnetic fields carrying currents in the same direction likely played a role in the generation of the X1.6 flare.
Active region NOAA 12192, though powerful, exhibited no eruptions. This intriguing case highlights the need for broader research. We should analyze the current evolution in both polarities of various flaring and non-flaring regions. By studying a larger sample of active regions in the future, we can gain a deeper understanding of how these currents behave and how their behavior in both polarities influences flare occurrence.


\begin{acks}
We thanks the reviewer for insightful comments. SDO is a mission of NASA's Living With a Star Program. S.K.T. gratefully acknowledges support by NASA HGI (80NSSC21K0520) and HSR (80NSSC23K0093) grants, and NSF AAG award (no. 2307505).
\end{acks}

{\bf Author contributions} Partha Chowdhury conceived the initial idea for this study, selected the active regions, and performed preliminary analyses. Ravindra B contributed to the data analysis. PC and RB co-wrote the manuscript. Sanjiv Tiwari provided valuable feedback and suggested several revisions.

{\bf Data Availability} The data used from HMI, AIA are freely available and downloaded from JSOC webpage.

{\bf Declarations \\
Competing interests} The authors declare no competing interests.  \\

{\bf Funding } S.K.T. gratefully acknowledges support by NASA HGI (80NSSC21K0520) and HSR (80NSSC23K0093) grants, and NSF AAG award (no. 2307505). \\

\bibliography{reference.bib}
\bibliographystyle{spr-mp-sola.bst}

\end{article}

\end{document}